\documentclass[conference]{IEEEtran}
\IEEEoverridecommandlockouts

\usepackage{cite}
\usepackage{amsmath,amssymb,amsfonts}
\usepackage{graphicx}
\usepackage{textcomp}
\usepackage{xcolor}
\usepackage{url}
\usepackage{booktabs}

\def\BibTeX{{\rm B\kern-.05em{\sc i\kern-.025em b}\kern-.08em
T\kern-.1667em\lower.7ex\hbox{E}\kern-.125emX}}

\begin{document}

\title{LLMCrater: Lifecycle-Aware FAIR Metadata Generation using Large Language Models}

\author{
\IEEEauthorblockN{Dani Termaat}
\IEEEauthorblockA{
University of Amsterdam\\
The Netherlands\\
d.k.termaat@uva.nl
}
\and
\IEEEauthorblockN{Nafiseh Soveizi}
\IEEEauthorblockA{
University of Amsterdam\\
The Netherlands\\
n.soveizi@uva.nl
}
\and
\IEEEauthorblockN{Zhiming Zhao}
\IEEEauthorblockA{
University of Amsterdam\\
The Netherlands\\
z.zhao@uva.nl
}

\and
\IEEEauthorblockN{Marios Avgeris}
\IEEEauthorblockA{
University of Amsterdam\\
The Netherlands\\
m.avgeris@uva.nl
}

}

\maketitle

\begin{abstract}

FAIR (Findable, Accessible, Interoperable, and Reusable) metadata is essential for the discovery, interoperability, and reuse of scientific research assets. However, creating and maintaining FAIR metadata remains largely manual, making the process time-consuming for heterogeneous research artifacts generated throughout the research lifecycle. Existing approaches primarily generate metadata at publication time, missing opportunities to capture contextual information as it becomes available. To address this limitation, we present \emph{LLMCrater}, a lifecycle-aware metadata generation framework that combines Large Language Models (LLMs) with stage-specific RO-Crate metadata profiles. The framework progressively enriches metadata across four research lifecycle stages (Design, Development, Deployment, and Execution \& Provenance) while remaining compatible with RO-Crate~1.1 and EOSC metadata recommendations. It automatically extracts metadata from heterogeneous artifacts, generates and validates machine-actionable RO-Crates, and supports publication to FAIR repositories and PID services (e.g., Zenodo). We demonstrate the approach using two representative use cases: a 5G experimentation environment within SLICES-RI and an experiment on GreenDIGIT's EcoJupyter platform. Results show that LLMCrater progressively enriches metadata throughout the research lifecycle and generates valid RO-Crates conforming to the RO-Crate~1.1 specification.\footnotemark
\end{abstract}

\footnotetext{Source code: \url{https://github.com/DT-UVA/LLMCrater-V2}}



\section{Introduction}
\vspace{-2mm}
Research infrastructures (RIs) continuously produce diverse digital assets, including datasets, software, workflows, notebooks, infrastructure configurations, and experimental outputs. Although these assets are typically published with FAIR metadata, the metadata itself evolves throughout the research lifecycle~\cite{wilkinson2016fair}. Project objectives are defined during design, software and workflows emerge during development, infrastructure information becomes available during deployment, while provenance and execution records are produced during experimentation. Capturing metadata only at publication therefore misses valuable contextual information that could support earlier sharing, collaboration, and progressive FAIRification.

Current FAIR publication workflows generate metadata only after a project or experiment has been completed, despite metadata naturally evolving throughout the research lifecycle. Consequently, contextual and provenance information may be lost, limiting progressive FAIRification, early sharing, and collaboration. At the same time, standards such as RO-Crate~\cite{rocrate} and EOSC metadata recommendations~\cite{eoscroadmap,eoscsemantic} provide a common foundation for interoperable, machine-actionable research objects, but creating compliant metadata remains labor-intensive.

To address these limitations, we present \emph{LLMCrater}, a lifecycle-aware metadata generation framework that combines Large Language Models (LLMs) with stage-specific RO-Crate metadata profiles. Rather than generating metadata only at publication time, LLMCrater progressively enriches metadata across four lifecycle phases (Design, Development, Deployment, and Execution \& Provenance). The framework automatically extracts metadata from heterogeneous research artifacts, generates and validates RO-Crates, and supports publication to repositories such as Zenodo while remaining compatible with RO-Crate~1.1 and EOSC metadata recommendations. We demonstrate the framework using the SLICES-RI 5G experimentation environment~\cite{slices} and GreenDIGIT's federated EcoJupyter platform~\cite{greendigit_d51}.


\vspace{-1mm}
\section{Lifecycle-Aware Metadata Generation}
\vspace{-1mm}
Instead of generating metadata only at publication time, LLMCrater progressively enriches RO-Crates throughout the research lifecycle. Four stage-specific metadata profiles define the mandatory and optional metadata expected for the Design, Development, Deployment, and Execution \& Provenance phases. At each stage, LLMCrater generates a new RO-Crate that extends its predecessor with newly available metadata rather than replacing it. A single research project is therefore represented by a sequence of progressively richer RO-Crates, each capturing the state of the research object at a specific lifecycle stage. All profiles remain compatible with RO-Crate~1.1 and EOSC metadata recommendations.



\begin{table}[!t]
\centering
\caption{Representative metadata captured across the four lifecycle phases.}
\label{tab:lifecycle_profiles}
\footnotesize
\begin{tabular}{p{0.17\linewidth}p{0.7\linewidth}}
\hline
\textbf{Phase} & \textbf{Representative metadata} \\
\hline
Design &
Title, description, objectives, scientific motivation, research questions, contributors, planned datasets, models, workflows, infrastructure, expected outputs, funding. \\
\hline
Development &
Source repositories, versions, commit IDs, programming languages, dependencies, workflow definitions, documentation, software licenses, contributor roles, I/O specifications. \\
\hline
Deployment &
Container images, VMs, cloud providers, hardware, operating systems, service endpoints, infrastructure configurations, environment variables, security settings. \\
\hline
Execution \& Provenance &
Input datasets and models, runtime parameters, execution context, timestamps, outputs, provenance relationships, derived assets, DOIs, citation metadata, FAIR assessment results, version lineage. \\
\hline
\end{tabular}
\end{table}

By capturing metadata incrementally as research progresses, the proposed lifecycle-aware model preserves contextual information throughout the complete research process and produces progressively richer, machine-actionable research objects that are ready for FAIR publication and reuse.
\vspace{-2mm}
\section{LLMCrater Architecture}
\vspace{-2mm}

\begin{figure}[!t]
    \centering
    \includegraphics[width=0.8\linewidth]{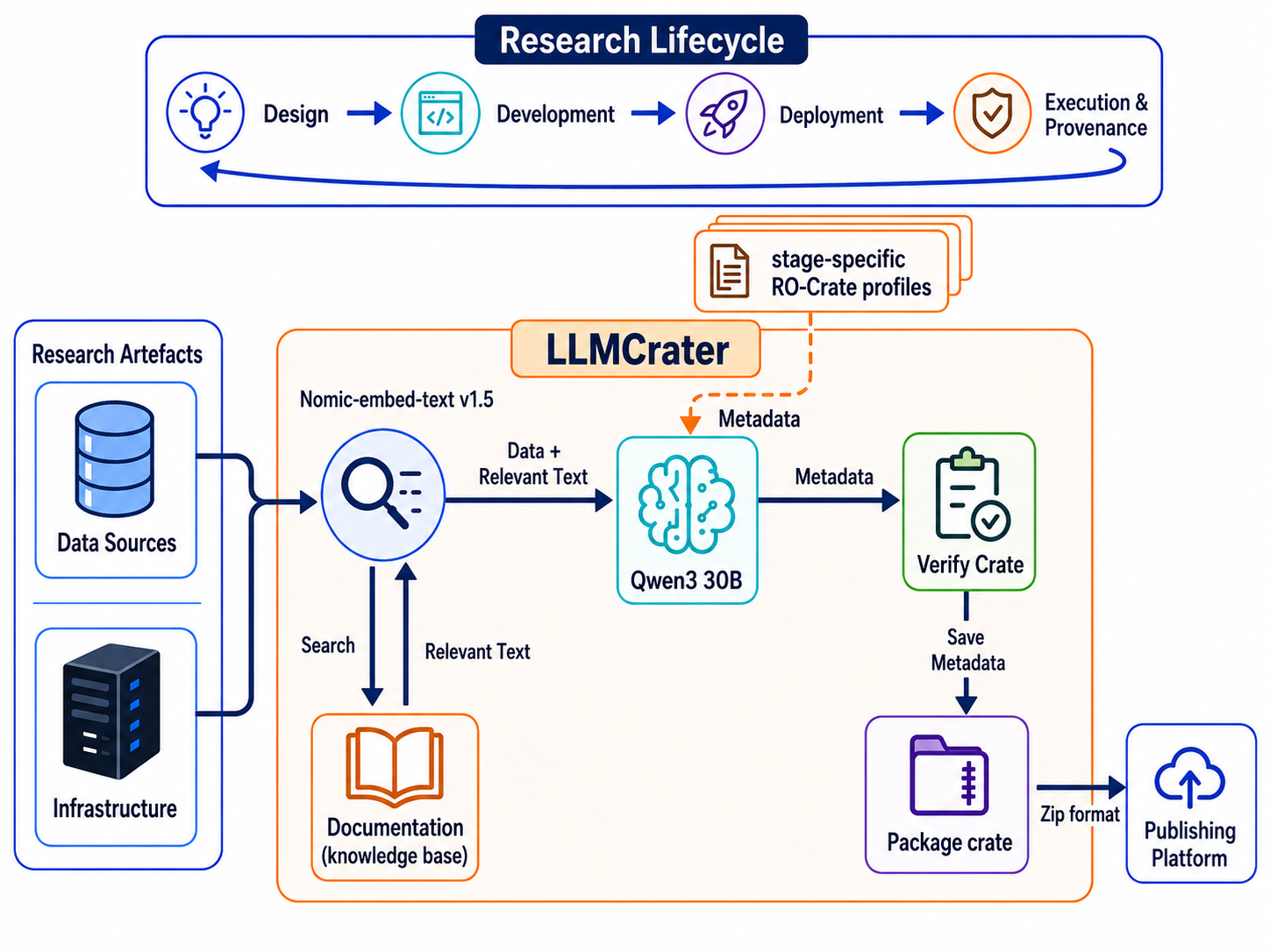}
    \caption{LLMCrater Architecture.}
    \label{fig:LLMCraterArch}
\end{figure}

Figure~\ref{fig:LLMCraterArch} illustrates the lifecycle-aware LLMCrater architecture. The framework automatically generates and enriches FAIR metadata for heterogeneous research artifacts, including datasets, software repositories, notebooks, workflow descriptions, infrastructure configurations, and documentation. Based on the current lifecycle phase (Design, Development, Deployment, or Execution \& Provenance), the corresponding stage-specific RO-Crate profile guides metadata generation. Although demonstrated using SLICES-RI and GreenDIGIT, the architecture is infrastructure-agnostic and can be adopted by other research infrastructures, such as LifeWatch~\cite{lifewatch}, that manage heterogeneous research artifacts throughout the research lifecycle.

\vspace{-2mm}
LLMCrater follows a Retrieval-Augmented Generation (RAG) workflow implemented using LangChain. Documentation describing the selected lifecycle profile together with the RO-Crate~1.1 specification is embedded using \textit{nomic-embed-text-v1.5} and indexed in a Chroma vector database. During generation, the retrieved guidance is combined with the research artifacts and supplied to the \textit{Qwen3-30B} model (executed locally through Ollama), enabling generation of metadata that conforms to both the selected lifecycle profile and the RO-Crate~1.1 specification. The generated JSON-LD metadata is validated using the RO-Crate Python library before being packaged together with the research assets as an RO-Crate. The resulting research object can optionally be published directly to Zenodo, where it receives a persistent DOI, enabling citation, sharing, and long-term accessibility.

\section{Experimental Evaluation}
The proposed framework was evaluated through a proof-of-concept using two heterogeneous use cases: (i) the SLICES-RI 5G experimentation environment and (ii) GreenDIGIT's EcoJupyter platform. The datasets were enhanced with synthetic properties to cover the complete metadata model. Together, the use cases comprise diverse research artifacts, including Jupyter notebooks, datasets, software repositories, workflow descriptions, infrastructure configurations, and documentation.

For each use case, LLMCrater extracted metadata from the available artifacts, selected the appropriate lifecycle-specific RO-Crate profile, generated machine-actionable JSON-LD metadata using a RAG-enhanced LLM pipeline, validated the resulting RO-Crate against the RO-Crate~1.1 specification, and packaged the research object for optional publication to Zenodo. The generated metadata included infrastructure- and experiment-level information, software dependencies, execution environments, datasets, workflow components, and provenance metadata. In both case studies, the generated RO-Crates passed validation and were successfully packaged as publication-ready research objects. Table~\ref{tab:results} summarizes the proof-of-concept results.

\begin{table}[t]
\centering
\caption{Growth of RO-Crate complexity across the research lifecycle.}
\label{tab:results}
\footnotesize
\begin{tabular}{p{0.4\linewidth}p{0.25\linewidth}p{0.15\linewidth}}
\toprule
\textbf{Lifecycle Stage} & \textbf{Entities \& Relationships} & \textbf{Metadata Types} \\ \hline
\midrule
Design        & 20  & 5  \\\hline
Development   & 41  & 8  \\\hline
Deployment    & 49  & 10 \\\hline
Execution \& Provenance    & 70  & 14 \\\hline
\bottomrule
\end{tabular}
\end{table}



\section{Acknowledgement}
\footnotesize{This work was partially supported by the European Commission through the Horizon Europe SLICES-IP project (Grant No. 10128769), by the Dutch Research Council (NWO) Large-Scale Research Infrastructures (LSRI) programme for the LTER-LIFE infrastructure (Grant No. 184.036.014), by LifeWatch ERIC, and by the European Union through the projects ENVRI-Hub Next (101131141), EVERSE (101129744), BlueCloud-2026 (101094227), OSCARS (101129751), and BMD (101181294).}

\end{document}